\newcommand{\gSize}{1}

\newcommand{\Graph}[2]{
\begin{figure}[hbt]
\centering
\includegraphics[width=\gSize\columnwidth]{#1}
\caption{#2}
\label{fig:#1}
\end{figure}
}

\newcommand{\EditA}[2]{#1}
\newcommand{\EditB}[2]{#1}

\documentclass[final,5p,times,twocolumn]{elsarticle}

\usepackage{amssymb}
\usepackage{amsmath}

\usepackage{lineno}

\usepackage{textcomp} 
\usepackage{color}
\journal{NIM A}

\begin{document}

\begin{frontmatter}



\title{Digital Data Acquisition For the Low Energy Neutron Detector Array (LENDA)}


\author{S. Lipschutz$^{1,2,3}$}
\author{R.G.T. Zegers$^{1,2,3}$}
\author{J. Hill$^{1,2,3}$}
\author{S. N. Liddick$^{2,4}$}
\author{S. Noji$^{2,3}$\corref{cor1}}
\author{C. J. Prokop$^{2,4}$\corref{c3or1}}
\author{M. Scott $^{1,2,3}$\corref{c3or1}}
\author{M. Solt$^{2,3,5}$\corref{cor22}}
\author{\\C. Sullivan$^{1,2,3}$\corref{cor31}}
\author{J. Tompkins$^{2}$\corref{cor13}}


\cortext[cor1]{Present Address: Research Center for Nuclear Physics (RCNP), Osaka University,
10-1 Mihogaoka, Ibaraki, Osaka, 567-0047, Japan\\
Electronic Address: noji@rcnp.osaka-u.ac.jp}

\cortext[cor22]{Present Address: Department of Physics, Stanford University, Stanford, CA, USA, Electronic Address: mrsolt@stanford.edu}

\address{1 Department of Physics and Astronomy, Michigan State University, East Lansing MI, USA \\
2 National Superconducting Cyclotron Laboratory, East Lansing MI, USA \\
3 Joint Institute for Nuclear Astrophysics Center for the Evolution of the Elements, Michigan State University, East Lansing, MI, USA\\
4 Department of Chemistry, Michigan State University, East Lansing MI, USA\\
5 Department of Physics, Oakland University, Rochester MI, USA
}

\begin{abstract}
A digital data acquisition system (DDAS) has been implemented for the Low Energy Neutron Detector Array (LENDA).  LENDA is an array of 24 BC-408 plastic-scintillator bars designed to measure low-energy neutrons with kinetic energies in the range of 100 keV to 10 MeV from (p,n)-type charge-exchange reactions.  Compared to the previous data acquisition (DAQ) system for LENDA, DDAS offers the possibility to lower the neutron detection threshold, increase the overall neutron-detection efficiency, decrease the dead time of the system, and allow for easy expansion of the array.  The system utilized in this work was XIA's Digital Gamma Finder Pixie-16 250 MHz digitizers.  A detector-limited timing resolution of 400 ps was achieved for a single LENDA bar.  Using DDAS, the neutron detection threshold of the system was reduced compared to the previous \EditA{analog}{analogue} system, now reaching below 100 keV.  The new DAQ system was successfully used in a recent charge-exchange experiment using the $^{16}$C(p,n) reaction at the National Superconducting Cyclotron Laboratory (NSCL).

\end{abstract}


\end{frontmatter}


\section{Introduction}
\label{}

The Low Energy Neutron Detector Array (LENDA) \cite{Perdikakis2012} is a time-of-flight neutron-detector system consisting of 24 BC-408 plastic-scintillator bars each with dimensions of  45 mm (width), 25 mm (depth) and 300 mm (height).  LENDA was developed for studying (p,n)-type charge-exchange (CE) reactions at intermediate energies ($\sim$100 MeV/u) in inverse kinematics with rare-isotope beams.  It is designed to detect neutrons with kinetic energies in the range of $\sim$100 keV - 10 MeV.  Studying CE reactions at these energies are of strong interest for investigating the isovector response of neutron-rich nuclei, including giant resonances \cite{Harakeh2001,Osterfeld1992,Fujita2011}.  In \EditB{particular the Gamow-Teller}{particular Gamow-Teller} (GT) giant resonance has important applications for astrophysical phenomena such as core-collapse supernova \cite{LMP2003}.

The (p,n) experiment in inverse kinematics involves the detection of low-energy neutrons in coincidence with fast residues in a magnetic spectrometer.  A commissioning experiment was performed with LENDA and the S800 Spectrograph \cite{Bazin2003} using the $^{56}$Ni(p,n) and $^{55}$Co(p,n) reactions at 110 MeV/u \cite{sasano2011,sasano2012}.  The neutron laboratory scattering angle and neutron kinetic energy determined from LENDA provided full event-by-event kinematic information.  The CE reaction product (e.g. $^{56}$Cu in the case of the $^{56}$Ni) or one of its decay products ($^{55}$Ni, $^{54}$Co) were detected in the S800 Spectrograph\EditB{}{,} where a tag on the charge-exchange channel was provided.  The implementation of LENDA for these experiments utilized \EditA{analog}{analogue} electronics\EditB{.  In}{; in} this work the development and instrumentation of LENDA using a digital data acquisition system (DDAS) is discussed.

In order to be an effective solution for CE studies, DDAS must preserve the advantageous properties that are expected from plastic scintillators (fast timing and high neutron detection efficiency \cite{Knoll2010}), while introducing improved features that were unavailable in the \EditA{analog}{analogue} system.  The main challenge for the implementation of DDAS is to retain good timing resolution (sub-nanosecond) from data that is only sampled every few nanoseconds. Separate from this challenge, DDAS offers the possibility to improve the performance of the detector in a variety of areas.  The neutron detection threshold can be lowered, which translates to an improved neutron detection efficiency across all neutron energies.  Further, DDAS has near zero dead time and can tolerate higher data rates.  \EditA{In the configuration for LENDA the dead time of an individual DDAS channel has been measured to be $\sim$400 ns.}{} Sections \ref{algorithms} through \ref{timming res section} will discuss the different approaches for sub-nanosecond timing with DDAS and the results obtained with LENDA. Sections \ref{energyMethods} and \ref{threshold} will discuss the signal-amplitude resolution and neutron detection threshold achieved with LENDA readout in DDAS.

\section{Digital Data Acquisition System}
The fundamental function of DDAS is to produce a digitized waveform, which is referred to as the trace.  Each signal from the detector system is sent through a Nyquist filter and into a high-frequency \EditA{analog}{analoque} to digital converter (ADC), where the analogue voltage is sampled to produce a digital waveform \cite{XIA2005}.



In this \EditB{work,}{work} LENDA was instrumented with XIA's Digital Gamma Finder Pixie-16 boards  \cite{XIA2005}.  For each of the 16 channels on the board, 250 \EditB{megasamples}{mega samples} per second (MSPS), 14-bit flash ADCs were used to perform the digitization \EditB{of}{on} the incoming detector signals.  After capturing the waveforms, 4 field-programmable gate arrays  (FPGAs) implement on-board trace-processing routines to extract both time and energy information for each pulse (these will be discussed in more detail in sections \ref{algorithms} and \ref{energyMethods}).  The Pixie-16 modules were placed in a PXI/PCI crate, which was controlled by a National Instruments MXI-4 crate controller connected with fiber-optic cable to a dedicated computer.  

\section{Digital Timing Algorithms}
\label{algorithms}
Unlike the \EditA{analog}{analogue} system where timing and energy measurements were performed \EditB{by}{on} two independent circuits, DDAS extracts this information simultaneously from the  trace.  For both time and energy this is accomplished through digital filtering algorithms, either on the board or in off-line analysis.  For a general description of digital filtering as it applies to energy extraction and high-precision spectroscopy, see e.g. Ref. \cite{Jordanov1994}.  Of interest to this work, is the effectiveness of different timing algorithms with the digital data obtained from LENDA and their ability to achieve LENDA's \EditA{previously measured resolution of approximately 420 ps.}{}


For the 250-MSPS digitizers (period of 4 ns between clock ticks) used in this work, some form of interpolation scheme is needed to achieve timing resolution below the sampling frequency.  This can be accomplished by selecting an appropriate function for the detector signal of interest and \EditB{fitting}{fit} that function to the waveform to extract a sub clock tick time for each pulse.  This method was successfully implemented for the plastic scintillator array VANDLE \cite{Paulauskas2014} where a detector-limited timing resolution of 0.6 ns was achieved over the full energy range using the fit function,
\begin{equation}
f(t)=\alpha e^{-(t-\phi)/\beta}(1-e^{-(t-\phi)^4/\gamma}).
\end{equation}

\noindent
\EditB{With the highest energy signals, a}{With highest energy signals a} timing resolution of 0.5 ns was achieved. The fitting method requires significant computational post processing and the creation of a library of appropriate functions for each detector signal.  An alternate choice, and the one utilized in this work, is to use a digital constant fraction discriminator (CFD) algorithm similar to the one implemented on the Pixie boards \cite{XIA2005,prokop2015}.  Higher-order interpolations of the CFD are done off line in software that are unavailable on the board.  The implementation for this work is as follows.  The trace is processed through a symmetric trapezoidal filter,
\begin{equation}
FF[i]=\sum\limits_{j=i-(T_{r}-1)}^{i}Tr[j] - \sum\limits_{j=i-(2T_{r}+T_{g}-1)}^{i-(T_{r}+T_{g})}Tr[j]
\end{equation}
where $T_r$ is referred to as the \EditB{rise}{raise} time, $T_g$ the gap time and $Tr$ the trace points.  The filtered trace is used to form a digital CFD,

\begin{equation}
CFD[i+D]=FF[i+D]-w~FF[i]
\end{equation}
where $D$ is a user specified delay and $w$ is a scaling factor \cite{XIA2005,Prokop2014}.  The units of $T_r$, $T_g$, and $D$ are in clock ticks and $w$ is a unitless scaling factor.    The sub clock-tick timing information is then determined from a linear interpolation of the CFD's zero crossing (this technique will be referred to as the \textit{Linear Algorithm}).  In Fig. \ref{fig:cfd_trace.pdf}, an example of this procedure is shown.  The filled circles are a trace for a gamma ray from a $^{22}$Na source generated by one photo multiplier tube (PMT) of a LENDA bar. The filled squares are its corresponding CFD signal with filter parameters $T_r=6$, $T_g=2$, $D=4$, $w=5/8$.  
\Graph{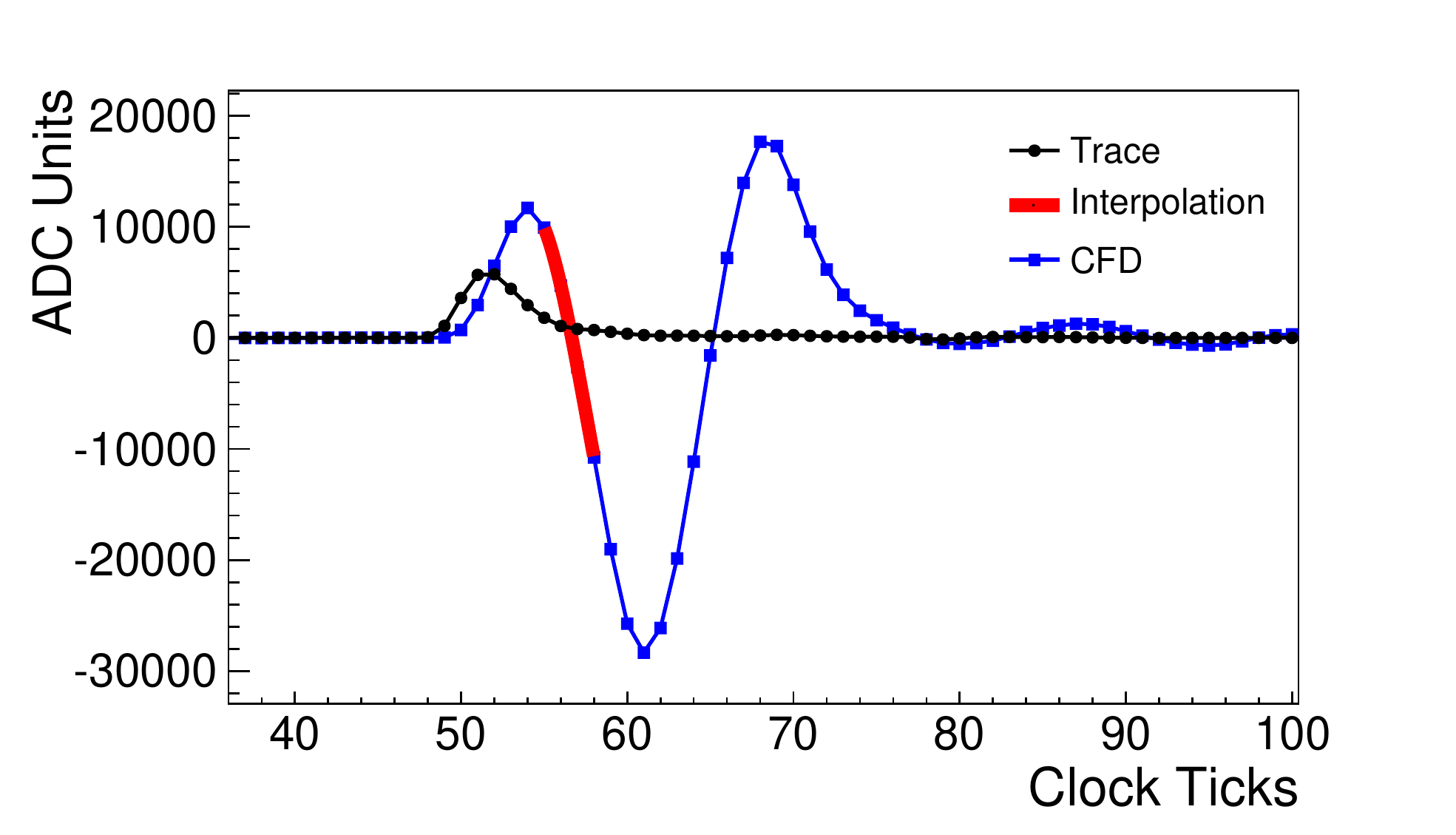}{(Color Online) Example of digital trace from LENDA (filled circles) with overlaid digital CFD (filled squares).  The thick line is the third-order interpolation function calculated for this event.}

\section{Intrinsic Timing Resolution of DDAS}
\label{Electronic Timing Resolution}
One of the significant challenges for high-precision timing with DDAS is the occurrence of non-linearities in the system's response to fast scintillator pulses \cite{Paulauskas2014,prokop2015}.  This behavior makes it difficult to rely on a simple linear interpolation of the CFD's zero crossing.  This non-linearity is maximized when the arrival time of a signal places its zero crossing in between two clock ticks, which for the 250 MHz digitizers used here, is a time difference of 2 ns.  To explore this behavior for LENDA, a $^{22}$Na source was placed on a LENDA bar where the signal from one PMT of the bar was split into DDAS with a 2 ns delay added to one copy of the signal.  Signals from both the 511 and 1275 keV gamma rays from $^{22}$Na were used in the following analysis.  The CFD algorithm described above was performed on each signal with the filter set $T_r=2$, $T_g=0$, $D=4$ and $w=1/8$.  This parameter set corresponds to the optimum choice for the \textit{Cubic Algorithm}, described below, and is one example of the linear-interpolation algorithm failing in the presence of a non-linear CFD signal.  Figure \ref{fig:3Timings.pdf} shows the time difference between these two signals after correcting for the dependence of the time difference on signal amplitude (a phenomenon known as ``walk").  \EditB{The walk correction was done by plotting the pulse height against the time difference and correcting for any observed correlation of the time difference on the pulse height.}{}  \EditB{The pulse heights in this data set ranged from approximately 4\% to 58\% of maximum ADC value.}{}  The linear interpolation with the above parameter set (labeled in the plot as ``Unoptimized Linear") introduces systematic biases and fails to achieve good timing resolution.

However, the shape of the digital CFD signal in the neighborhood of the zero crossing is a function of the CFD filter parameters.  By performing an optimization of those parameters significant improvement can be achieved in the performance of the linear-interpolation method.  As the parameters for the CFD algorithm are discrete, a brute force approach was used for the optimization.  Each trace in the data was processed with every possible filter combination in the ranges $T_{r}=1-8$, $T_{g}=0-3$, $D=1-8$, $w=1/8-7/8$ ($T_{r}$, $T_{g}$, $D$ were limited to integer values, $w$ was limited to steps of 1/8).  The parameter set that achieved the best resolution was taken as the optimum set.  The result of that optimization ($T_r=6$, $T_g=2$, $D=4$ and $w=5/8$) is shown in Fig. \ref{fig:3Timings.pdf}, where it is labeled as ``Optimized Linear".  Significant improvement is made in the linear method after the optimization, however as Fig. \ref{fig:3Timings.pdf} shows there is still some systematic bias in the distribution.

\Graph{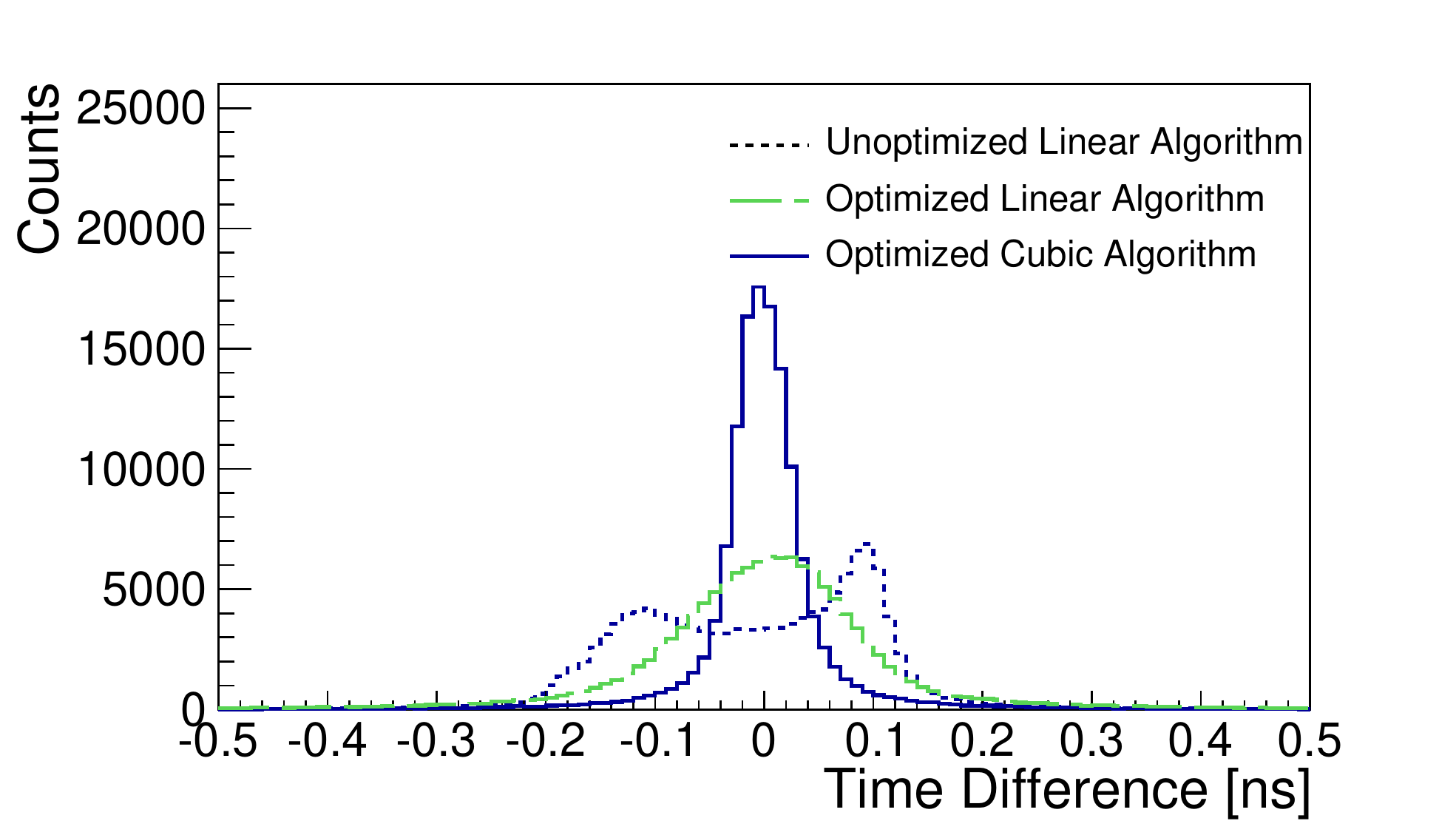}{(Color Online) Time-difference spectra from a split-signal electronics test.  The time differences from the 3 different timing algorithms have been shifted to zero for comparison.}

Further improvements can be made through the use of a higher-order interpolation of the  CFD signal's zero crossing.  Here the two points before the zero crossing and the two points after are used to calculate a third-order polynomial interpolating function.  An example of this is shown in Fig. \ref{fig:cfd_trace.pdf} as a thick red line.  Given the clock ticks and corresponding ADC values, the coefficient for the interpolating function are found by inverting the following equation,
\begin{equation}
\begin{bmatrix}
    (x_{1})^3       & (x_{1})^2 & (x_{1})^1 & (x_{1})^0 \\
    (x_{2})^3       & (x_{2})^2 & (x_{2})^1 & (x_{2})^0 \\
    (x_{3})^3       & (x_{3})^2 & (x_{3})^1 & (x_{3})^0 \\
    (x_{4})^3       & (x_{4})^2 & (x_{4})^1 & (x_{4})^0 \\
\end{bmatrix}
\begin{bmatrix}
    c_{3}   \\
    c_{2}   \\
    c_{1}     \\
    c_{0}     \\
\end{bmatrix}
=
\begin{bmatrix}
    y_{1}   \\
    y_{2}   \\
    y_{3}     \\
    y_{4}     \\
\end{bmatrix}
\end{equation} 
where $x_{1,2}$ and $y_{1,2}$ are the two points before the zero crossing and $x_{3,4}$ and $y_{3,4}$ are the two points after.  The same optimization procedure as described above was used for this method (referred to as the \textit{Cubic Algorithm}).  The result of this optimization ($T_r=2$, $T_g=0$, $D=4$ and $w=1/8$) applied to the same data set as above is also shown in Fig. \ref{fig:3Timings.pdf}, labeled as ``Optimized Cubic".  It is clear that the resolution has improved substantially using the cubic interpolation scheme and any systematic biases in the time difference spectrum have been removed.  

The intrinsic resolution results using the four points and a third-order polynomial were much better than the expected resolution of the detector (\EditA{$\sim$420.}{$\sim$500} ps)  Therefore, higher-order or more complicated interpolating schemes were not pursued in this work.


\EditB{Fig.}{Figure} \ref{fig:otherElectronicRes.pdf} shows the full-width at half-maximum (FWHM) electronic timing resolution of DDAS from the same data as shown in Fig. \ref{fig:3Timings.pdf}.  The width of the time difference distribution is plotted as a function of pulse height for both the optimized cubic- and linear-interpolation algorithms.  Figure \ref{fig:otherElectronicRes.pdf} shows that the \textit{Cubic Algorithm} performs significantly better at low pulse amplitudes, while at an appreciable fraction of the ADCs range both techniques provide electronic resolution well below the expected performance of a LENDA bar. 

\Graph{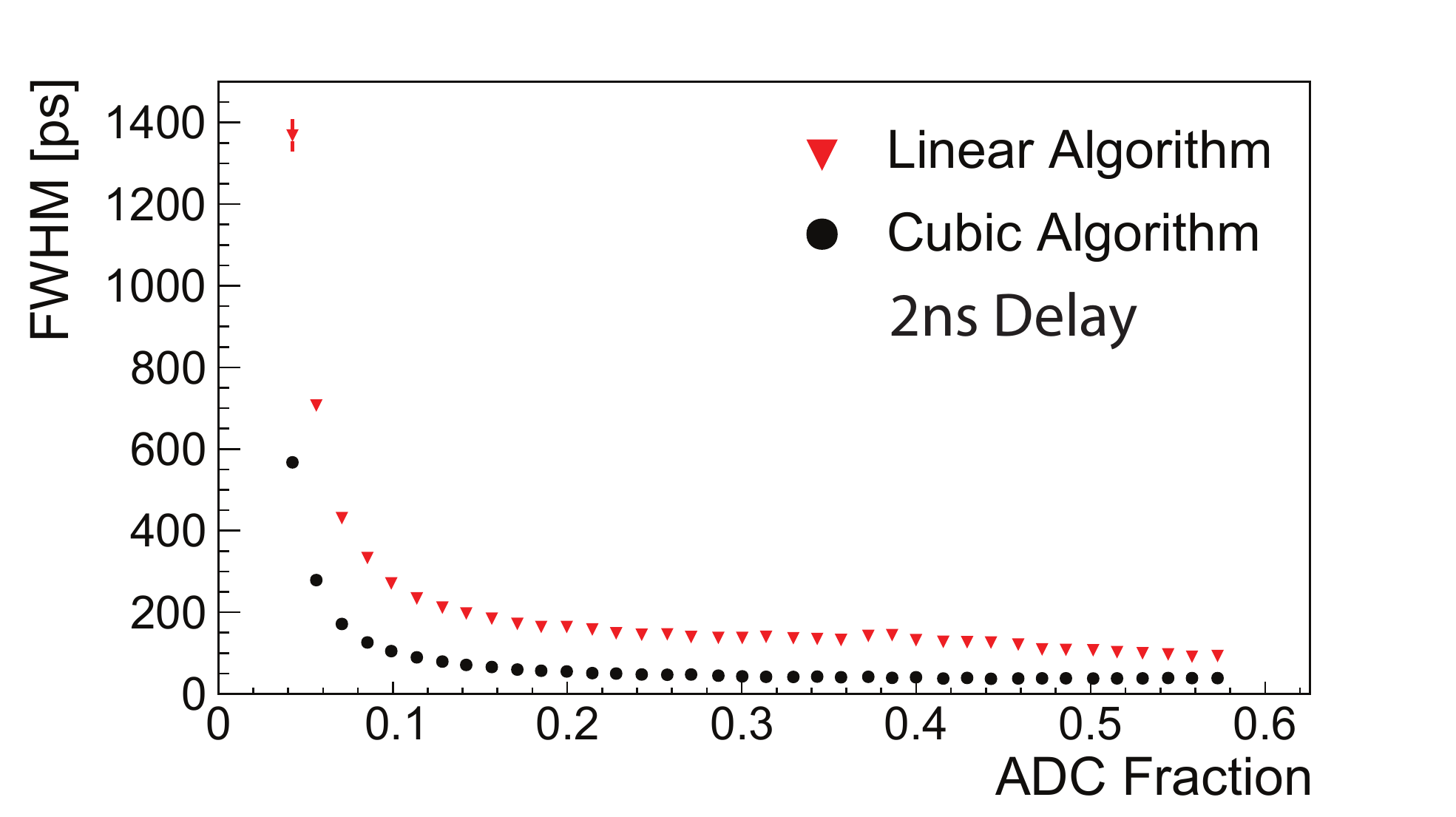}{(Color Online) Electronic timing resolution of DDAS as measured with a split signal from a LENDA bar.  The upside down triangles utilize a linear-interpolation method while the filled circles use a cubic-interpolation method.  See text for details.}

\section{Timing Resolution of LENDA with DDAS}
\label{timming res section}
The timing resolution of LENDA with DDAS was measured using the two correlated 511 keV gamma rays generated after the positron decay of $^{22}$Na.  Two LENDA bars were placed on opposing sides of the $^{22}$Na source at a distance of 1 m. Events were recorded in a 4-way coincidence between the PMTs of the two bars.  The time difference between the two bars was calculated as:
\begin{equation}
T_{\text{diff}}=\frac{1}{2}(T_{\text{1,top}}+T_{\text{1,bottom}}) -\frac{1}{2}(T_{\text{2,top}}+T_{\text{2,bottom}})
\end{equation}

\noindent
where the numerical subscripts refer to bar 1 and bar 2.  The width of the $T_{\text{diff}}$ distribution, corrected for walk, and divided by $\sqrt{2}$, is taken as the timing resolution of a single LENDA bar.  In the same way as above, the parameters for this data set were optimized for the best resolution, yielding $T_r=2$, $T_g=2$, $D=4$ and $w=1/8$ for the linear algorithm and $T_r=2$, $T_g=0$, $D=2$ and $w=1/8$ for the cubic algorithm.  \EditB{Fig.}{Figure} \ref{fig:Linear_Cubic_opt_Compare.pdf} shows the measured timing resolution in FWHM of one LENDA bar as function of pulse amplitude \EditA{and energy in keV$_{\text{ee}}$}{} for the linear and cubic timing algorithms discussed above.  Unlike the data presented in section \ref{Electronic Timing Resolution}, the signals in this measurement arrive randomly across all time differences, instead of always at a time difference of 2 ns.  Because of the intrinsic resolution of the detector, the difference in achieved resolution between the cubic and linear algorithms is less than above.  However, an improvement when using the cubic algorithm is still achieved, reaching about 400 ps at the highest \EditA{energies}{ADC values}.  \EditA{The saturation of the resolution at approximately 400 ps is a result of the intrinsic properties of the detector, not from the electronics.}{}

\Graph{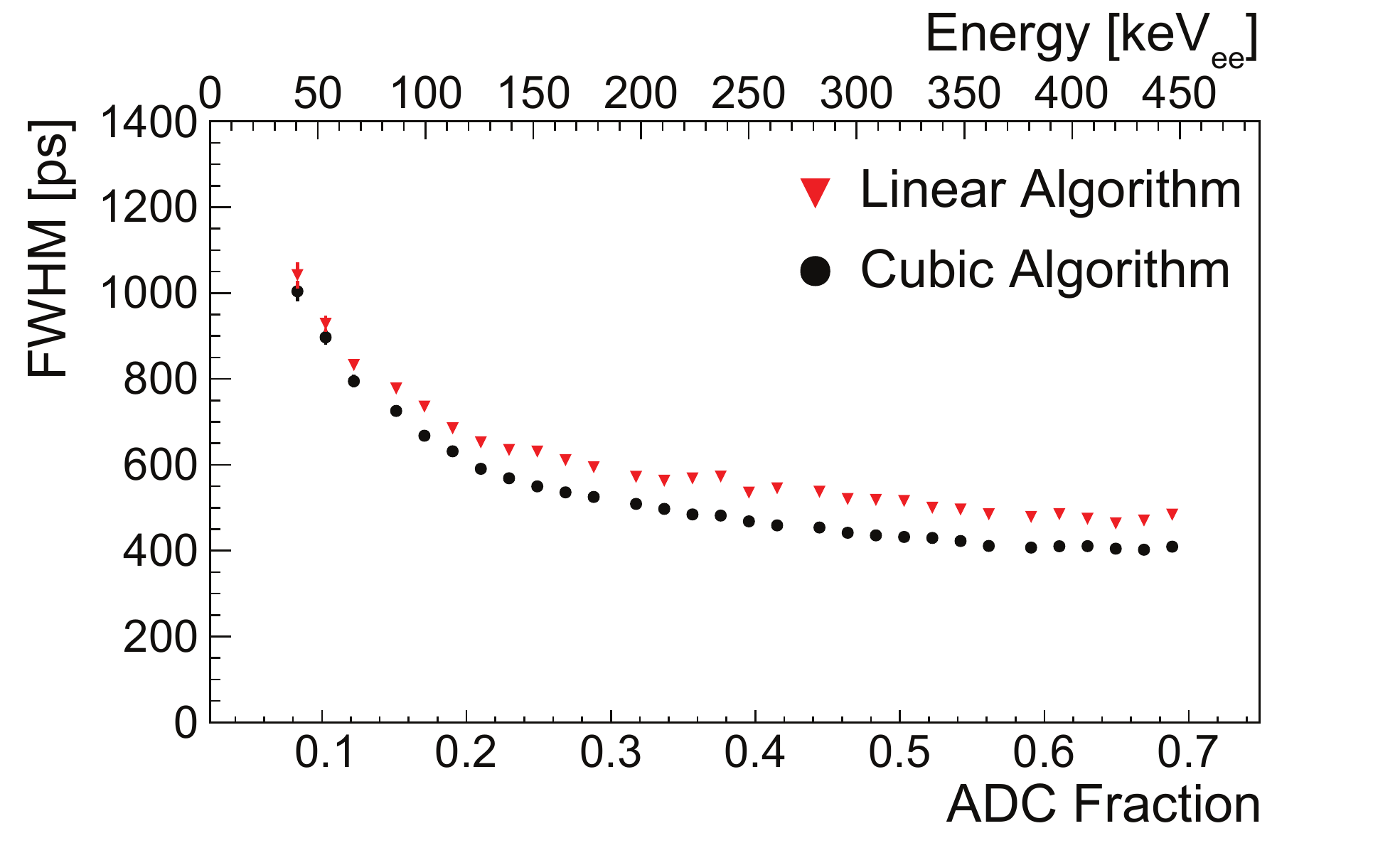}{(Color Online) Measured timing resolution for LENDA readout with DDAS.  Upside down triangles use a linear-interpolation method while the filled circles use a cubic-interpolation method.}


\EditB{When neutrons interact with the scintillator they can deposit any energy between 0 and their full kinetic energy into the bar.  Consequently, the timing resolution depends on the deposited energy (as shown in Fig. \ref{fig:Linear_Cubic_opt_Compare.pdf}) and not strongly on the neutron's kinetic energy.  On average the timing resolution for fast neutrons is slightly better than for low-energy neutrons because a larger fraction of events have a higher energy deposit.  However, low-energy neutrons will have a longer time-of-flight and therefore the somewhat reduced timing resolution has a lower impact on the energy resolution compared to fast neutrons.  As explained in Ref. \cite{Perdikakis2012}, an average timing resolution of approximately 0.5 ns across the range of pulse amplitudes leads to an almost constant neutron energy resolution of 5-6\%, which remains true when using DDAS.}{}

\section{Energy Extraction Methods}
\label{energyMethods}
There are many algorithms for extracting energy from digitized data, especially for high-precision spectroscopy (see, for example, Ref. \cite{Tan2004}).  In the case of a plastic scintillator array, where signal-amplitude resolution is expected to be limited, relatively straight forward methods can be utilized.  With the waveform saved, the energy of the event can be easily determined from the maximum trace value or, as is done here, by integrating the digitized pulse.  In Fig. \ref{fig:energywindownew.pdf} a trace from one PMT in LENDA is shown, taken with a $^{22}$Na source.  The beginning  24 clock ticks of the trace is used to determine the average baseline of the signal, while a window around the maximum point in the trace is used to find the pulse's amplitude.

\Graph{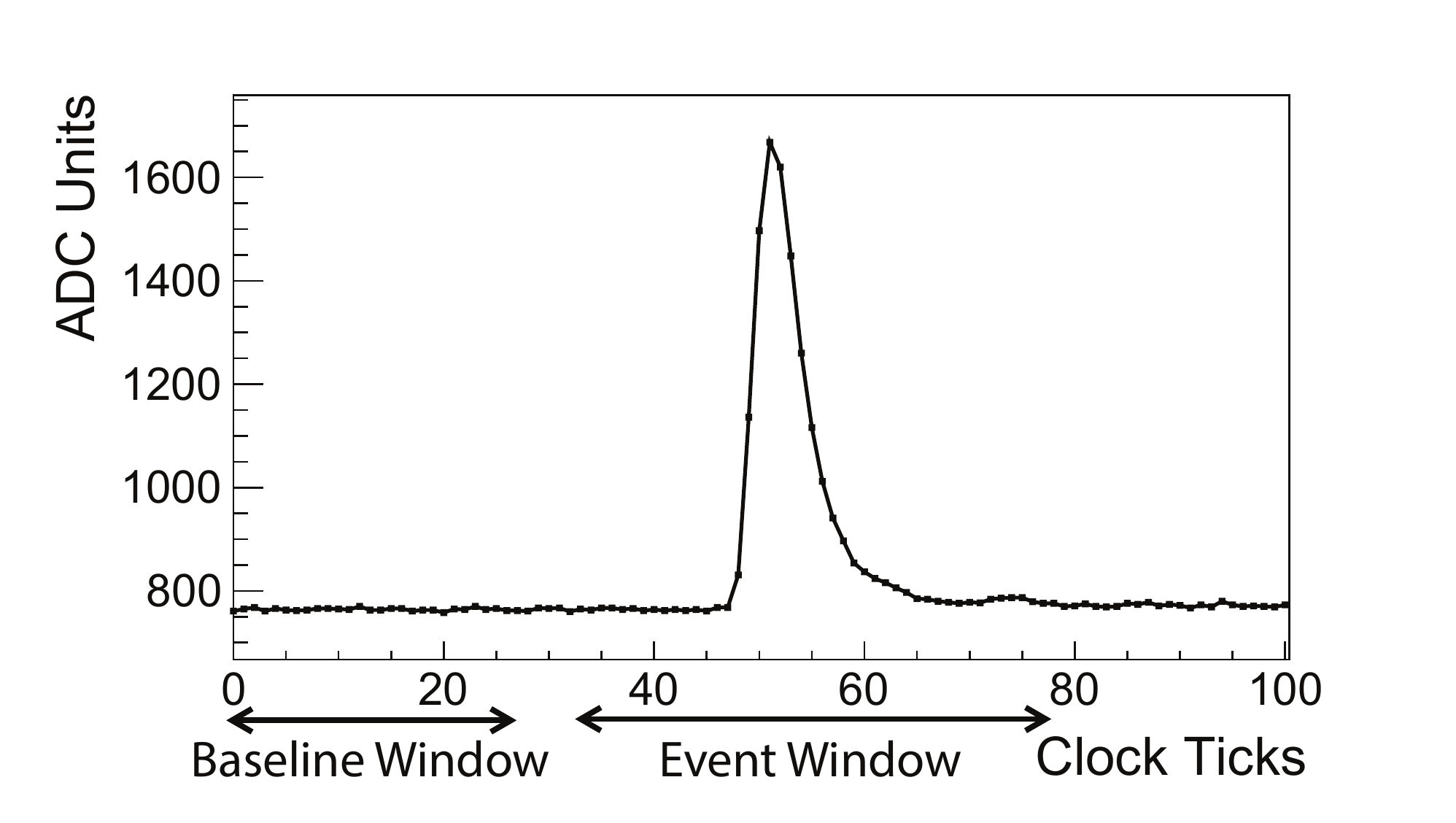}{A trace from a PMT of LENDA from a $^{22}$Na source.  The first integration region is used to determine the average baseline of the signal, while the second integrates the pulse and determines the energy of the event.}

To determine the resolution, an energy spectrum for $^{22}$Na was measured using a LENDA bar.  The source was placed in the center of the bar, and the energy of the event was taken as the geometric average of the top and bottom PMT's pulse integration values,
\begin{equation}
E_{Bar}=\sqrt{E_{T}*E_{B}}.
\end{equation}

The top panel in Fig. \ref{fig:Na22AndAm241.pdf} shows the calibrated energy spectrum in keV$_{\text{ee}}$ obtained from this measurement.  The two Compton edges from the 511 keV and 1274.5 keV gamma rays are seen at 341 and 1062 $\text{keV}_{\text{ee}}$, respectively.  The Klein-Nishina formula convolved with a Gaussian \cite{Kudomi1999} was used to fit the two Compton edges in order to extract a FWHM resolution of 23.5\% at 341 keV$_{\text{ee}}$ and 16\% at 1062 keV$_{\text{ee}}$.  Further, it is found that the position of the Compton edge is at 63.2 $\pm$ 1.9 \% of the maximum yield of the Compton spectrum, where the error is determined from the fitting of the Compton edges.

The bottom panel in Fig. \ref{fig:Na22AndAm241.pdf} shows an energy spectrum from a similar measurement using a $^{241}$Am source.  Here the photo peaks from the low-energy x-rays at 26.3 and 59.5 $\text{keV}_{\text{ee}}$ can be seen.  The photo peaks were fit with Gaussians to obtain a FWHM energy resolution of approximately 42.3\%.

\Graph{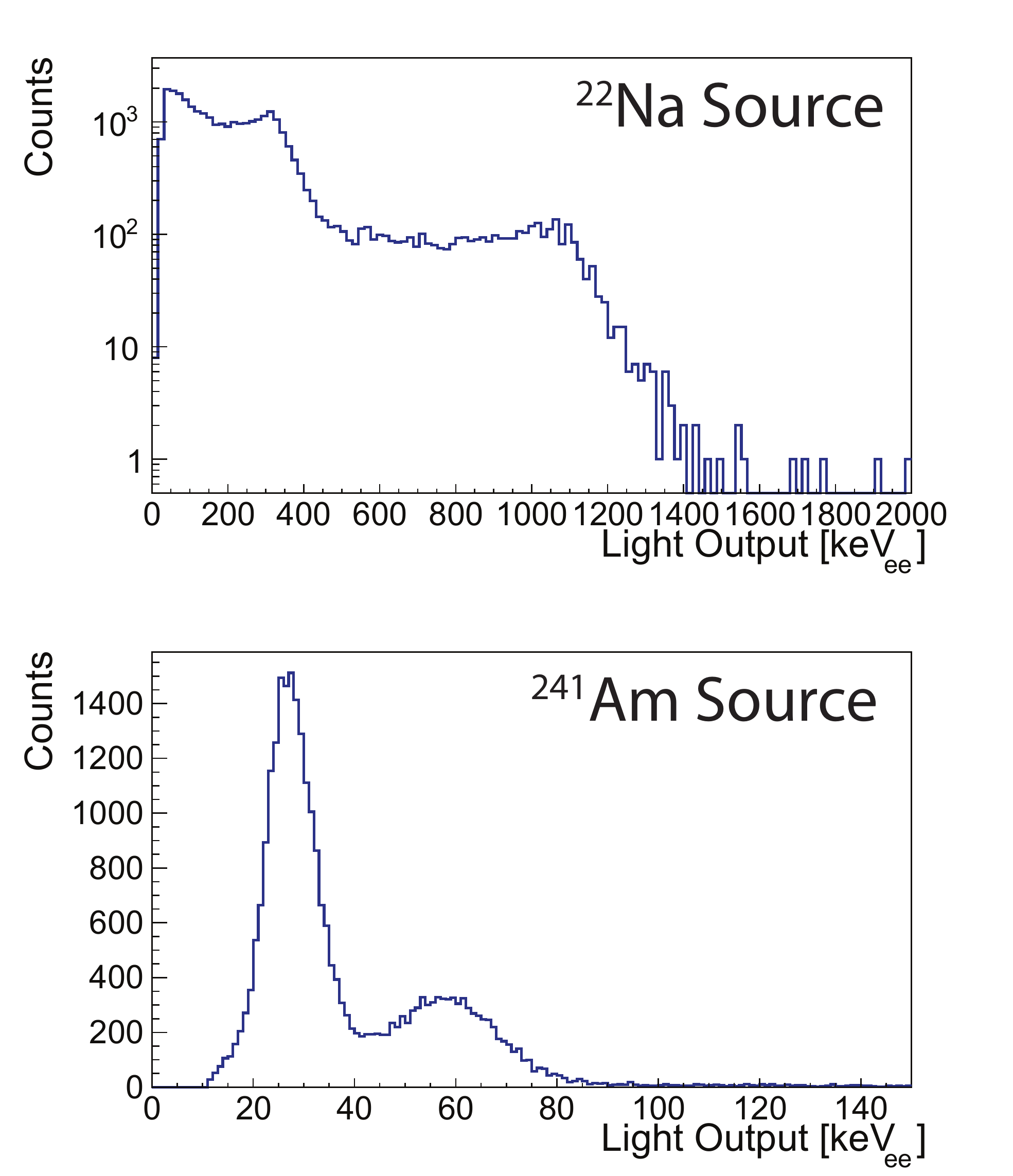}{Top panel shows a calibrated energy spectrum from $^{22}$Na seen in a LENDA bar.  Bottom panel shows a calibrated energy spectrum from a $^{241}$Am source.}

\section{Threshold Estimation}
\label{threshold}
The neutron detection threshold for LENDA with DDAS can be estimated using the gamma sources $^{22}$Na and $^{241}$Am.  \EditB{Fig.}{Figure} \ref{fig:threshGraph.pdf} shows the relationship between pulse integration in ADC units and energy in keV$_{\text{ee}}$ determined from the positions of the two Compton edges and the two photo peaks seen in Fig. \ref{fig:Na22AndAm241.pdf}.  The red line is a linear fit to the points.  \EditB{Fig.}{Figure} \ref{fig:threshGraph.pdf} indicates that the light output of LENDA readout in DDAS is proportional to energy deposit and that a threshold of approximately 10 keV$_{\text{ee}}$ is achieved.

\Graph{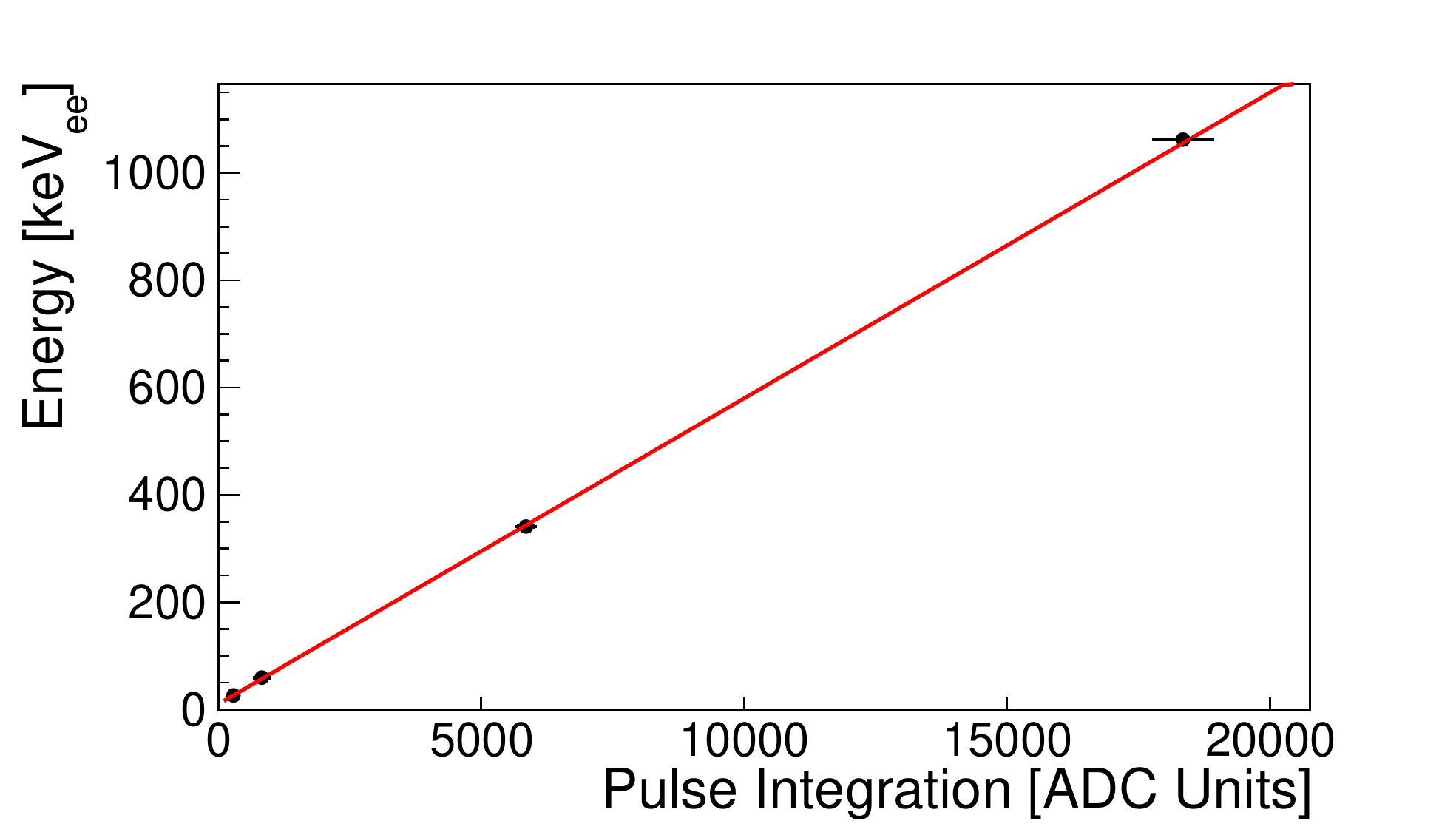}{Light output from LENDA (keV$_{\text{ee}}$) and ADC integration value measured with $^{22}$Na and $^{241}$Am sources.  The lower two points correspond to the two photo peaks in $^{241}$Am at  26.3 and 59.5 keV.  The higher two points are the Compton edges from $^{22}$Na at 341 and 1062 keV$_{\text{ee}}$.}

The correspondence between keV$_{\text{ee}}$ and neutron kinetic energy in keV can be measured using a  $^{252}$Cf fission source. This measurement was performed with an EJ-301 liquid scintillator placed next to the $^{252}$CF source, which was positioned 1 m from a LENDA bar.  The liquid scintillator served as the start signal for a neutron time-of-flight (TOF) measurement with the LENDA bar, where a 3-way coincidence between the two PMTs of the LENDA bar and the PMT of the liquid scintillator was required.  Background measurements were performed using copper shadow bars in a similar setup to that described in Ref. \cite{Perdikakis2012}. \EditB{Fig.}{Figure} \ref{fig:LO_TN.pdf} shows a result of this measurement, where light output in keV$_{\text{ee}}$ from neutrons in LENDA vs their kinetic energy determined from time-of-flight is presented.  For high neutron energies ($>$3 MeV) the relationship between maximum light output (full energy of the neutron deposited in the bar) and kinetic energy is linear.  The relationship approximately follows,

\begin{subequations}
\begin{equation}
\EditB{L_{max}(E)}{L(E)}=518.1~T_{n} -499.5,~~T_{n}>\text{3 MeV}
\end{equation}
where \EditB{$L_{max}(E)$}{$L(E)$} is the \EditB{maximum}{ } light output in keV$_{\text{ee}}$ and $T_{n}$ is the kinetic energy of the neutron in MeV.  This relationship is non-linear at low neutron energies where it follows,
\begin{equation}
\EditB{L_{max}(E)}{L(E)}=18.53+95.08~T_{n}+81.58~T_{n}^2,~~T_{n}<\text{3 MeV}.
\end{equation}
\end{subequations}

\Graph{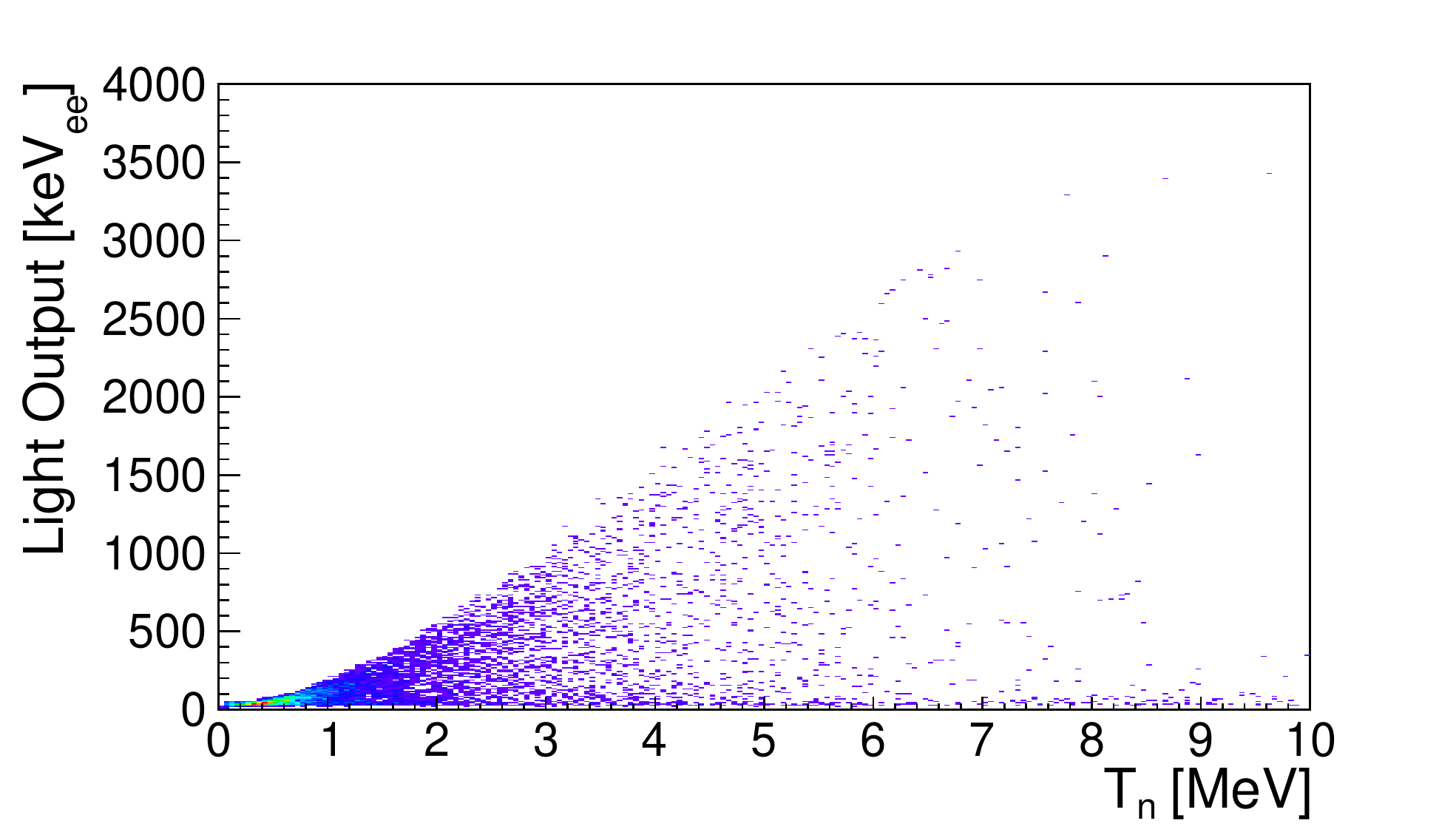}{Scatter plot of light output against neutron energy from a $^{252}$Cf source measured with LENDA and a EJ-301 liquid scintillator as the timing reference.} 

In order to clearly see the behavior for near-threshold neutrons a background-subtracted energy spectrum for the  data was produced and is shown in Fig. \ref{fig:CF_Plot.pdf}.  The black data points are the background-subtracted data in 200 keV bins.  The inset shows the background-subtracted energy spectrum for energies below 1 MeV in bins of 50 keV.  It is clear that neutrons of less than 100 keV are seen at a light output threshold of 10 keV$_{\text{ee}}$ indicating that the neutron detection threshold for LENDA is below 100 keV.  The neutron detection efficiency of LENDA was determined in a similar way as for the measurements with the \EditA{analog}{analogue} DAQ (see ref. \cite{Perdikakis2012}).  For a the same light output threshold DDAS produced neutron detection efficiency values consistent with those seen in the \EditA{analog}{analogue} DAQ.

\Graph{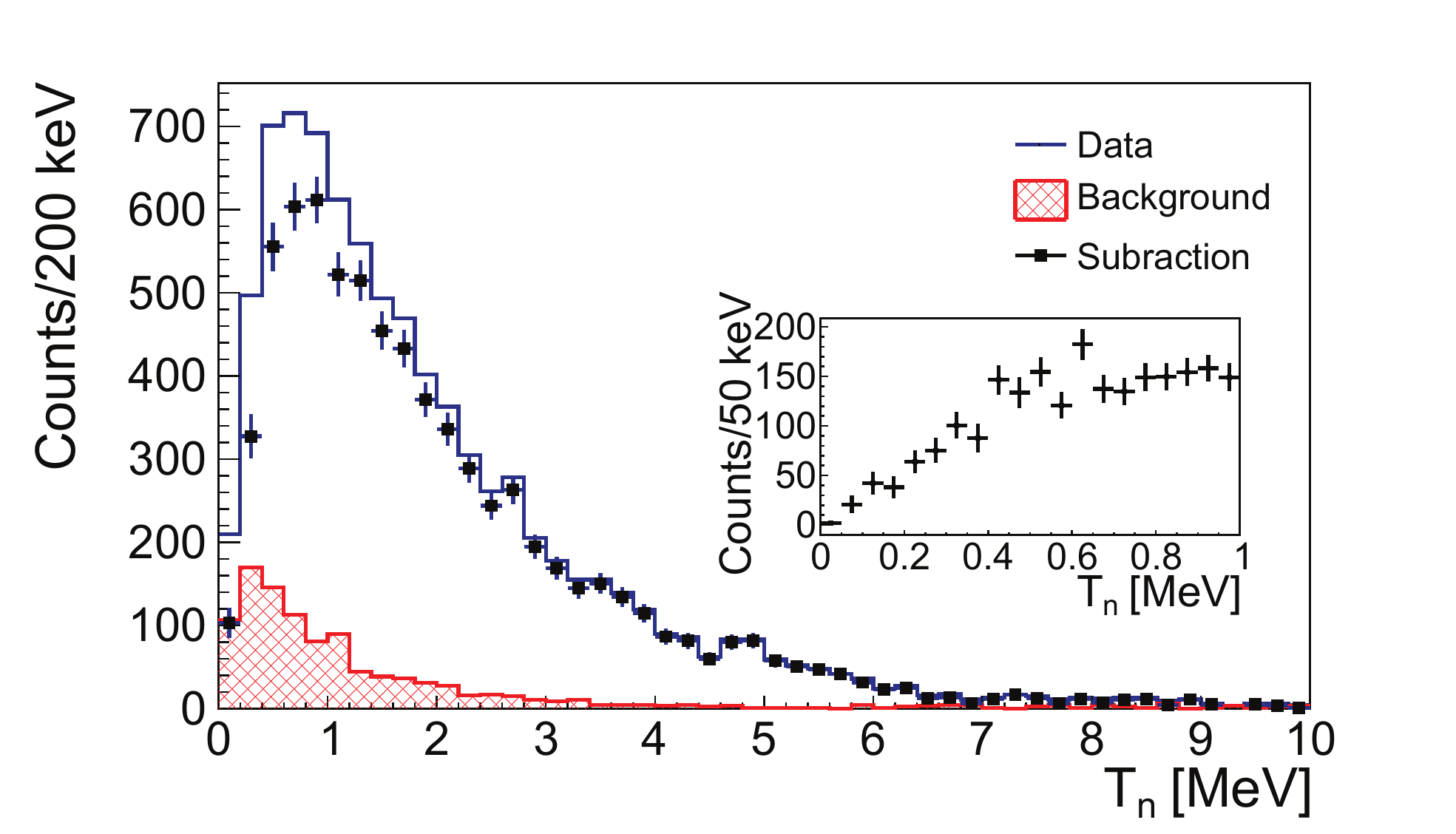}{(Color Online) Neutron energy spectrum from a $^{252}$Cf source, measured with LENDA.  The histogram shows the data, red hatched area is the result from a shadow-bar background measurement.  Black squares are the background-subtracted data.  Inset shows the low-energy tail of the background-subtracted distribution.}
\section{Experimental Uses}
LENDA with DDAS was successfully utilized in combination with the S800 Spectrograph in a recent charge-exchange experiment performed at the National Superconducting Cyclotron Laboratory.  A beam of $^{16}$C at 100 MeV/u was impinged on a liquid hydrogen target to study the $^{16}$C(p,n) reaction.  Beam-like fragments of $^{16}$N were detected in the S800 Spectrograph in coincidence with the neutrons in LENDA.  \EditA{The threshold settings for LENDA remained the same in this experimental run as they were in the data set presented above.}{} \EditA{The analog}{analogue} DAQ of the S800 was synchronized with the clock from DDAS, allowing for time-stamped event building to be done in software.  A 50 MHz clock was taken out of the DDAS crate, downscaled to 12.5 MHz and sent into a JTEC XLM72 module in a VME crate of the S800's DAQ.  For each event, the scaler for the clock signal was latched and readout into the data stream, allowing for a common time stamp between the two systems.  This was an important step towards future experiments involving the S800 and other digital systems.     
\section{Conclusion}
A new digital data acquisition system was successfully implemented with the Low Energy Neutron Detector Array at the NSCL.  Techniques for high precision timing with DDAS and LENDA were developed and it was shown that the achievable electronic timing resolution was far below the expected resolution of the detector.  Using these techniques a timing resolution of approximately 400 ps was achieved.  The neutron detection threshold for LENDA under DDAS was also investigated and found to be below 100 keV.  These properties of the system indicate that it performs the same or better as the previous \EditA{analog}{analogue} system. 




\section{Acknowledgments}
\noindent
This work was supported by the U.S. NSF Grant No. PHY-1430152 (Joint Institute for Nuclear Astrophysics Center for the Evolution of the Elements), U.S. Department of Energy National Nuclear Security Administration under Award Number DE-NA0000979 through the Nuclear Science and Security Consortium, and by the NSF under Cooperative agreement No. PHY-11-02511.

\bibliographystyle{elsarticle-num} 
\bibliography{MyBib}





\end{document}